# Itinerant Antiferromagnetism in FeMnP$_{0.8}$Si$_{0.2}$ Single Crystals

B.C. Sales, M. A. Susner, B. S. Conner, J. Q. Yan, and A. F. May
Materials Science and Technology Division Oak Ridge National Laboratory

(Dated: June 30, 2015)

Itinerant Antiferromagnetism in FeMnP$_{0.8}$Si$_{0.2}$ Single Crystals

B. C. Sales[1], M. A. Susner[1], B. S. Conner[1], J. Q. Yan[1,2] and A. F. May[1]

[1]Materials Science and Technology Division, Oak Ridge National Laboratory, Oak Ridge, Tennessee 37831
[2]Department of Materials Science and Engineering, University of Tennessee, Knoxville, TN 37996

**Abstract**

Compounds based on the Fe$_2$P structure have continued to attract interest because of the interplay between itinerant and localized magnetism in a non-centrosymmetric crystal structure, and because of the recent developments of these materials for magnetocaloric applications. Here we report the growth and characterization of mm size single crystals of FeMnP$_{0.8}$Si$_{0.2}$. Single crystal x-ray diffraction, magnetization, resistivity, Hall and heat capacity data are reported. Surprisingly, the crystals exhibit itinerant antiferromagnetic order below 158 K with no hint of ferromagnetic behavior in the magnetization curves and with the spins ordered primarily in the *ab* plane. The room temperature resistivity is close to the Ioffe-Regel limit for a metal. Single crystal x-ray diffraction indicates a strong preference for Mn to occupy the larger pyramidal 3g site. The cation site preference in the as-grown crystals and the antiferromagnetism are not changed after high temperature anneals and a rapid quench to room temperature.

I. INTRODUCTION

The magnetic properties of hexagonal $Fe_2P$ and related alloys have been heavily investigated for over forty years[1–7] and are intriguing because the observed behaviors have been described by the itinerant magnetism model of Moriya.[8] The $Fe_2P$ structure ($P\bar{6}2m$) has two distinct metal sites, $Fe_I$ and $Fe_{II}$. The smaller tetrahedral $Fe_I$ site (3f) has 4 P neighbors and the larger pyramidal $Fe_{II}$ site (3g) is coordinated by 5 P atoms. The pure compound exhibits a first order ferromagnetic transition below 210 K with magnetic moments on the $Fe_I$ and $Fe_{II}$ sites of 0.92 and 1.70 $\mu_B$, respectively.[9] The magnetism of $Fe_2P$ is extremely sensitive to pressure[10] and small concentrations of chemical dopants such as Mn, Si and B[3,11,12] can completely suppress ferromagnetism[3,10] or even triple the Curie temperature![11,12] The large difference between the two iron moments in $Fe_2P$ suggests a mixture of itinerant and more localized magnetism, which is even more evident in Mn doped alloys.[13] The recent interest in these compounds arises from two directions. First, the $Fe_2P$ crystal structure lacks inversion symmetry and hence these materials may exhibit complex canted spin structures and spin vortices such as "skyrmions"[14] that can be manipulated with a magnetic field or an electrical current.[15] The complex spin structures are due to the effects of spin-orbit coupling as parameterized by the Dzyaloshinskii-Moriya (DM) interaction,[16,17] which is finite in non-centrosymmetric crystal structures. Second, various alloy compositions $(Fe,Mn)_2(P,As,Si,B)$ with the $Fe_2P$ structure show great promise as magnetocaloric materials for refrigeration.[13,18–20] Because of the extreme sensitivity of these materials to relatively small changes in composition and the difficultly in preparing single-phase polycrystalline samples, the reported properties of many of these materials vary significantly due to the presence of up to 15% of impurity phases.[21] While some of these polycrystalline materials may be suitable for magnetocaloric applications, obtaining a fundamental microscopic understanding of the magnetism and the magneto-elastic coupling becomes difficult without high purity single crystal samples. This provided the motivation for the present study. Although there have been a few single crystal investigations of the Fe$_2$P type materials,[1–3,6] to our knowledge we have prepared the first single crystals with compositions similar to those proposed for magnetocaloric applications.[13,20]

## II. SYNTHESIS AND EXPERIMENTAL METHODS

Single crystals of Fe$_{2-x}$Mn$_x$P$_{1-y}$Si$_y$ are grown out of a molten Sn flux. High purity Fe and Mn pieces (Alfa Aesar, 99.98%), P chunks (Alfa Aesar, 99.999%), freshly ground Si powder (Alfa Aesar, 99.999%), and Sn shot (Alfa Aesar, 99.999%) are loaded into a 10 cc alumina crucible in a typical molar ratio Fe:Mn:P:Si:Sn of 2:2:1:1:30. The alumina crucible containing all of the elements and a second "catch" crucible with quartz wool are sealed inside an evacuated silica tube. The mixture is heated to 1125 °C held for 24h and then cooled to 900 °C at 1.5 °C/h. The ampoule is then removed from the furnace and the excess molten Sn is centrifuged into the catch crucible. In most cases small needlelike crystals with the Fe$_2$P structure weighing a few micrograms are obtained. These crystals are suitable for magnetization measurements and a few other experiments[22] but are too small for electrical transport and heat capacity measurements. The Fe to Mn ratio in the crystals could be controlled and is close to the starting ratio as long as the Fe/Mn ratio is between 1 and 1.8. For Fe/Mn values much less than 1, Mn$_2$P crystals tend to grow, while for Fe/Mn values larger than 1.8, Fe$_5$Si$_3$ crystals dominate. The P/Si ratio is more difficult to control, and the maximum amount of Si in crystals with the hexagonal Fe$_2$P phase is close to 0.3 (i.e. P$_{0.7}$Si$_{0.3}$) for the Fe/Mn ratios near 1.8, and about 0.2 for Fe/Mn ratios near 1. The limited control of the Si concentration in this type of growth is likely related to the smaller solubility of Si in molten Sn relative to the other elements.[23] For compositions near FeMnP$_{0.8}$Si$_{0.2}$ substantially larger crystals weighing a few mg are obtained (Fig. 1c). These larger crystals frequently have Sn inclusions and sometimes a Sn core running along the length of the crystal which had to be ground away before measurements. This is extremely important for electrical measurements because of the high electrical conductivity of Sn relative to FeMnP$_{0.8}$Si$_{0.2}$ and the giant magnetoresistance of Sn at low temperatures.[24] For the resistivity and Hall measurements, the crystals are polished to a 45 micron thick plate. Electrical leads are attached to the crystals using 0.05 mm Pt wires and Epo-Tek H20E silver epoxy. The Hall data are taken using the standard four point geometry[25] at fixed temperatures and with H varied from -6 to +6 Tesla in 1 Tesla steps. The Hall resistivity, $\rho_H$, is obtained from the odd in H part of the transverse resistivity $\rho_H = (\rho_{xy}(H) - \rho_{xy}(-H))/2$. The chemical compositions of the crystals are determined using a Hitachi 3000 Tabletop scanning electron microscope with a Bruker Quantax 70 energy dispersive X-ray spectroscopy (EDS) system. From EDS measurements there is no evidence of Sn incorporation into the $Fe_2P$ structure.

Single crystal x-ray diffractions patterns are collected on a Rigaku single crystal diffractometer using Mo K$\alpha$ radiation. Each single crystal diffraction experiment utilized ~150 unique peak reflections for refinements to determine crystal structure. Magnetic data are collected using a Magnetic Property Measurement System from Quantum Design (QD). Heat capacity, resistivity, and Hall data are collected using a QD Physical Property Measuring System.

## III. RESULTS AND DISCUSSION

The structure of $FeMnP_{0.8}Si_{0.2}$ is shown is figure 1, along with a photograph of a crystal. From the single crystal structure refinement, the R1 factors are smallest with Mn fully occupying the larger pyramidal 3g site and Fe the tetrahedral 3f site. The refinements are either unstable or yield an increase in the R1 factor if the metal sites are switched or randomized. No site preference for Si and P are evident in the refinements using the same technique. If the Mn and Fe sites are completely ordered, as indicated by the aforementioned results, the structure can be regarded as layered with alternating Mn and Fe layers stacked along the *c* axis (Fig. 1 b). The structure is refined above (220 K) and below (110 K) the antiferromagnet transition temperature $T_N$ = 158 K (Tables 1 and 2). For these crystals there is no significant change in the lattice parameters or the volume upon cooling or warming through the magnetic transition, unlike pure Fe$_2$P[1] and many of the similar ferromagnetic magnetocaloric compositions (e.g. FeMnP$_{0.67}$Si$_{0.33}$)[26].

The magnetic properties of the crystals are shown in Figs. 2-4. The susceptibility data, taken with an applied field of 1 Tesla (Fig. 2), indicates a magnetic transition at T$_N$ ≈ 158 K. Heating or cooling through the magnetic transition at 2 K/min, results in about 1 K of hysteresis, which is small relative to many of the ferromagnetic compositions.[1,13,19,20] The magnetization curves (M vs. H) are linear at all temperatures (Fig 3) with no evidence of a ferromagnetic intercept. Taken together Figs. 2 and 4 indicate antiferromagnetic order below 158 K with the spins oriented primarily in the ab plane.[27] The magnetically ordered state is not simple, however, as is evident from the magnetic susceptibility, $\chi$, data above T$_N$. Above T$_N$, there is a maximum in the susceptibility at 295 K. From 400-700 K the decrease in $\chi$ is linear in T ($\chi$ = 7.8 x 10$^{-5}$ – 5.96 x 10$^{-8}$ T cm$^3$/g), with no evidence of Curie-Weiss behavior. Qualitatively similar results were obtained for Fe$_2$P crystals.[6] For Fe$_2$P, neutron scattering data indicated substantial short-range order up to T=3T$_c$ ≈ 654 K. Only above 700 K was Curie-Weiss behavior recovered.[6]

Recent studies of polycrystalline samples with a nominal composition of FeMnP$_{0.75}$Si$_{0.25}$ found that the magnetic behavior was very different depending on whether the samples were quenched or slow-cooled.[28–30] The quenched samples exhibited ferromagnetism and strong hysteresis with a Curie temperature near 250 K, while samples cooled more slowly showed an antiferromagnetic component with a Neel temperature near 150 K and very little hysteresis. Based on Mossbauer, magnetization, crystallographic data, and theory these authors concluded that the magnetic ground state depended on the degree of Fe/Mn cation order. Complete occupancy of the 3g site with Mn favored antiferromagnetic order. Although, the polycrystalline samples had small amounts of other impurity phases[28,30], and even the "purest" phases had a mixture of ferromagnetic and antiferromagnetic order, the itinerant antiferromagnetic state is likely the same we observe in single crystals of FeMnP$_{0.8}$Si$_{0.2}$. Powder neutron diffraction from polycrystalline samples with a significant fraction of the antiferromagnetic phase indicated that the antiferromagnetism is incommensurate, with a wave vector $q_x$ = 0.361. The proposed magnetic structure[30] is non-collinear with the spins lying in the *ab* plane. This magnetic structure is consistent with our single crystal magnetic susceptibility data (Fig. 2) and the heat capacity data as discussed below.

Motivated by the very recent polycrystalline studies[28–30] on nominal $FeMnP_{0.75}Si_{25}$, single crystals of $FeMnP_{0.8}Si_{0.2}$ were wrapped in Ta foil to reduce oxidation, sealed under vacuum in a silica tube and then heated to 1060 C for 16 h followed by a rapid quench into water. For the quenched crystals no hint of ferromagnetism is observed in M vs H measurements at 5 K, but the Neel temperature was lowered slightly to about 152 K. These results suggest that the ordered arrangement of the Mn and Fe cations is likely the thermodynamically stable configuration for this composition.

Heat capacity data from the same crystal used for the magnetic measurements are shown in Figs. 5 and 6. In the temperature range from 2-200 K, the only phase transition evident is the magnetic transition at $T_N \approx$ 158 K. The shape and magnitude of the heat capacity peak are similar to that observed in other materials that exhibit a spin density wave transition (SDW) associated with an itinerant magnetic state.[31,32] The entropy associated with the transition at 158 K is estimated to be $\Delta S \approx$ 0.21 J/K-per transition metal, which is only about 3% of Rln2. The small entropy change is consistent with an itinerant antiferromagnet, although it is also likely that much of the entropy in the spin system is removed at higher temperatures due to short-range magnetic order. A standard Debye fit of the low temperature heat capacity (T<8 K) to $C_p \approx \gamma T + \beta T^3$, yields $\gamma$ = 8.5 mJ/$K^2$-mole atoms, and a Debye temperature of $\Theta_D$ = 360 K. This assumes that the magnetic contribution to $C_p$ below 8 K can be neglected. These values are similar to those reported for $Fe_2P$ of 8mj/$K^2$-mole atoms and $\Theta_D$ = 420 K[33].

Resistivity data, $\rho_c$, from a thinned crystal plate ($\approx$ 0.045 x 0.35 x 3 mm) are shown in Fig. 7 with the current along the *c* axis. The resistivity is weakly temperature dependent, but metallic at all temperatures. The magnitude of the resistivity is relatively high for a normal metal, but is typical for a semimetal or a heavily doped semiconductor. The sharp drop in the resistivity below $T_N$ is seen in most magnetic systems and is normally attributed to less magnetic scattering of the carriers in the magnetically ordered state. The magnitude of $\rho_c$ at 300 K is similar to that reported for a Fe$_2$P crystal ($\approx$ 300 $\mu\Omega$-cm)[1]. For temperatures below $T_N$, however, the decrease in $\rho_c$ in only about 40%, presumably because of the added chemical disorder relative to pure Fe$_2$P. The apparent carrier concentration, $n_{app}$, is determined from the Hall coefficient, $R_H$, using a simple one band model ($R_H \approx 1/ne$). This is plausible since all of the Hall resistivity data are linear in magnetic field up to at least 6 Tesla (not shown). Above $T_N$ the dominant carriers are holes and $n_{app} \approx 2.6 \times 10^{22}$ holes/cm$^3$. Just below $T_N$ there is an abrupt increase in $n_{app}$, followed by a more gradual net increase in $n_{app}$ down to 2 K. This behavior is unexpected since for most itinerant antiferromagnets part of the Fermi surface develops a gap below $T_N$, which should result in fewer carriers not more. An apparent increase in the density of states below a magnetic transition is non-intuitive. One possible explanation for the Hall data below $T_N$ is that it is related to the Anomalous Hall Effect (AHE) often seen in ferromagnets.[34] The increase in $n_{app}$ below $T_N$ looks roughly like the temperature dependence of the local staggered magnetization in an antiferromagnet, but further research is needed to test this hypothesis. The transverse magnetoresistance is small at all temperatures ($\approx$ 0.04% at 6T), but is negative above $T_N$ and positive below (not shown). A positive magnetoresistance is normal for a metal, and the negative magnetoresistance above $T_N$ is normally associated with the proximity to ferromagnetism[35]. For FeMnP$_{0.8}$Si$_{0.2}$, the change in sign of the magnetoresistance at $T_N$ is at least an indicator of a change in the electronic structure at $T_N$.

As noted above, a room temperature resistivity of 250 $\mu\Omega$ cm is large for a metallic system, and may be close to the Ioffe-Regel limit, $\rho_{sat}$, where the mean free path of the carriers approaches the interatomic distance, $a$.[36] A simple model was proposed by Gurvitch[36] for $\rho_{sat}= 1.29 \times 10^{18}/n^{2/3}a$ μΩ-cm, where n is the carrier concentration per cm$^3$, and the interatomic distance, *a,* is in Å. At room temperature using a value of *a* = 3 Å this gives a value of $\rho_{sat}$ of 490 μΩ-cm for FeMnP$_{0.8}$Si$_{0.2}$, which is within a factor of 2 of the measured value at room temperature. The tendency of the resistivity to saturate at higher temperatures may reflect an approach to the Ioffe-Regel limit. We note that the data previously reported for the high temperature resistivity of Fe$_2$P single crystals shows very little temperature dependence above 300 K.[1]

## IV. SUMMARY AND CONCLUSIONS

In summary, mm-sized single crystals of FeMnP$_{0.8}$Si$_{0.2}$ are grown from a Sn flux. To our knowledge these are the first sizable single crystals grown with compositions close to those proposed for magnetocaloric applications.[20] The crystals are characterized with a variety of techniques including single crystal X-ray diffraction, magnetization, magnetic susceptibility, heat capacity, Hall and resistivity. Single crystal X-ray diffraction indicates a strong preference for Mn to occupy the larger pyramidal 3g site in the Fe$_2$P structure. Surprisingly, the crystals are antiferromagnetic below T$_N$≈158 K with the spins oriented primarily in the *ab* plane. Unlike many of the ferromagnetic magnetocaloric compositions[20], the antiferromagnetic transition is very weakly first order with only about 1 K hysteresis with heating and cooling rates of 2 K/min. From heat capacity data, the change in magnetic entropy near T$_N$ is rather small (about 3% of Rln2), which suggests that the magnetic transition is of an itinerant SDW type. This is supported also by very recent powder neutron diffraction from an impure polycrystalline sample with a significant fraction of this antiferromagnetic phase. The antiferromagnetic phase was found to be incommensurate with a wave vector $q_x$ = 0.361.[30] Single crystal magnetic susceptibility data above room temperature also suggests significant magnetic short range order similar to what is observed in pure Fe$_2$P.[6] Based on Hall data above T$_N$, the magnitude of the resistivity of FeMnP$_{0.8}$Si$_{0.2}$ (and likely that of $Fe_2P$) approaches the Ioffe-Regel limit for a metal. Below $T_N$, the increase in the apparent carrier concentration may be related to an anomalous Hall effect[37].

**ACKNOWLEDGEMENTS**

This work was supported by the U. S. Department of Energy, Office of Science, Basic Energy Sciences, Materials Sciences and Engineering Division. BSC acknowledges support from the Critical Materials Institute, an Energy Innovation Hub, funded by the U.S. Department of Energy, Office of Energy Efficiency and Renewable Energy, Advanced Manufacturing Office.

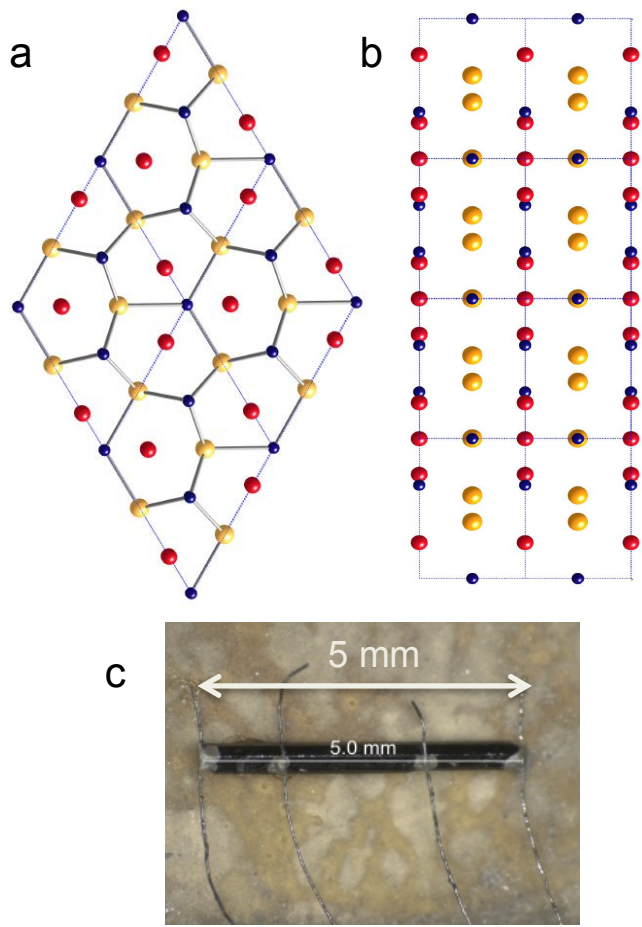

Figure 1: Crystal structure of FeMnP$_{0.8}$Si$_{0.2}$. View along the (001) direction (a), and view along (110) direction (b) The small P (Si) atoms are blue, the medium Fe atoms are red and the largest Mn atoms are yellow. The structure is drawn with Mn fully occupying the pyramidal (3g) site and Fe the (3f) tetrahedral site, consistent with the single crystal refinement results shown in Table I. A 5 mm long single crystal of FeMnP$_{0.8}$Si$_{0.2}$ with electrical leads attached.

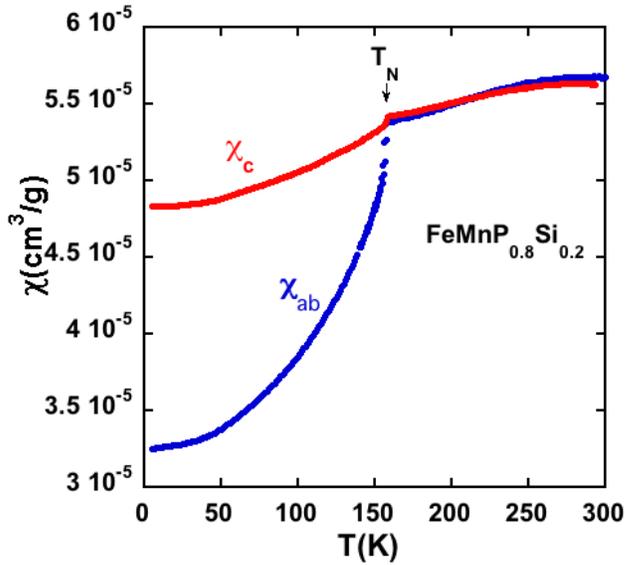

Fig 2: Magnetic susceptibility versus temperature for a FeMnP$_{0.8}$Si$_{0.2}$ single crystal with a magnetic field of 1 Tesla applied parallel or perpendicular to the *c* axis. The susceptibility data indicates antiferromagnetic order with the spins oriented primarily in the ab plane.

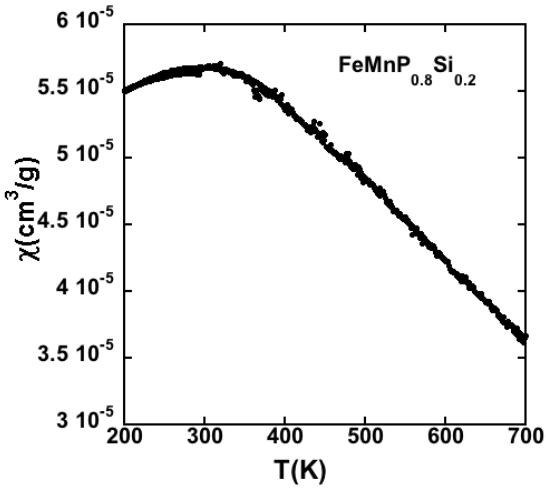

Fig 3: Magnetic susceptibility versus temperature above T$_N$≈ 158 K with an applied magnetic field of 1 Tesla. Between 400 and 700 K the decrease in $\chi$(T) is linear, which is unusual and indicates complex magnetic behavior.

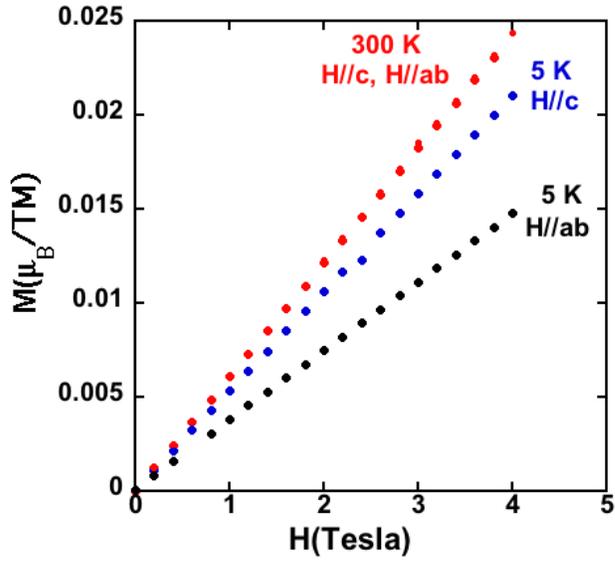

Fig. 4: Magnetization versus magnetic field. No evidence of ferromagnetism is observed.

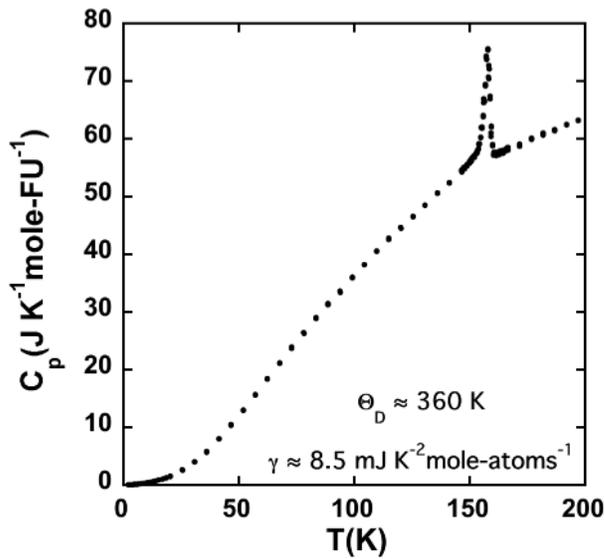

Fig. 5: Heat capacity versus temperature at H = 0T. No significant change in the data is observed for H = 8T.

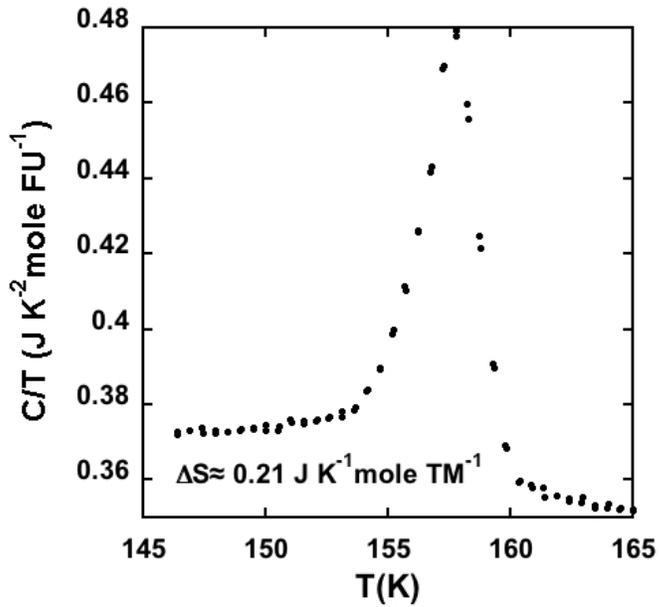

Fig. 6: Heat capacity divided by T versus T near the Neel temperature at H = 0T. No significant change in the data is observed for H = 8T. The entropy change is about 3.6% of Rln(2) per mole of transition metal.

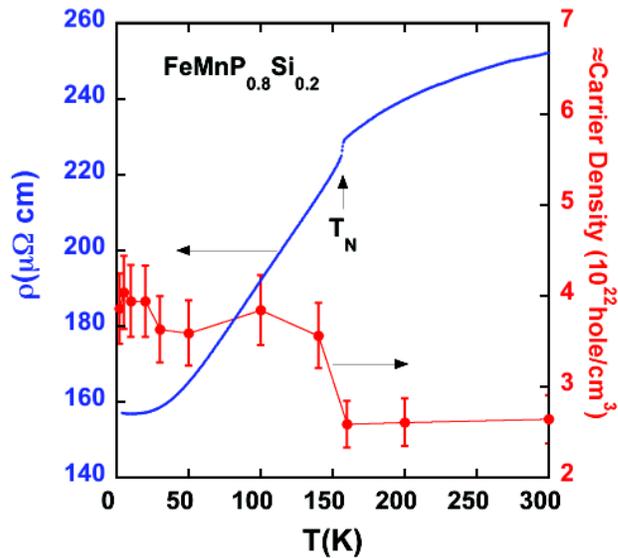

Fig 7: Resistivity versus temperature with I//c (left axis) Apparent carrier concentration versus temperature (right axis).

Table 1. 230 K- Structural Parameters for FeMnP$_{0.8}$Si$_{0.2}$ – Space Group $P\bar{6}2m$ (189)

| Lattice Constants | Site | Wycoff Position | Atom Coordinates | | | Occ. | Occ. f. |
|---|---|---|---|---|---|---|---|
| | | | x | y | z | | |
| a = b = 5.9681(38) Å | Fe(1) | 3f | 0.253932 | 0.00000 | 1.00000 | 0.25 | 1.000 |
| c = 3.4966(22) Å | Mn(2) | 3g | 0.589899 | 0.00000 | 0.50000 | 0.25 | 1.000 |
| $\alpha = \beta = 90°$  $\gamma = 120°$ | P(1) | 2c | 0.333333 | 0.66667 | 0.00000 | 0.166667 | 0.80725 |
| | Si(1) | 2c | 0.333333 | 0.66667 | 0.00000 | 0.166667 | 0.19275 |
| | P(2) | 1b | 0.00000 | 0.00000 | 0.50000 | 0.083333 | 0.80725 |
| | Si(2) | 1b | 0.00000 | 0.00000 | 0.50000 | 0.083333 | 0.19275 |

| Reliability Factors | $R_1$ | $wR_2$ | $R_{int}$ | GooF |
|---|---|---|---|---|
| | 0.0311 | 0.0600 | 0.0581 | 0.977 |

Table 2. 110 K -Structural Parameters for FeMnP$_{0.8}$Si$_{0.2}$ –Space Group $P\bar{6}2m$.

| Lattice Constants | Site | Wycoff Position | Atom Coordinates | | | Occ. | Occ. f. |
|---|---|---|---|---|---|---|---|
| | | | x | y | z | | |
| a = b = 5.9638(43) Å | Fe(1) | 3f | 0.254149 | 0.000000 | 0.00000 | 0.25 | 1.000 |
| c = 3.4903(25) Å | Mn(2) | 3g | 0.589859 | 0.000000 | 0.50000 | 0.25 | 1.000 |
| $\alpha = \beta = 90°$  $\gamma = 120°$ | P(1) | 2c | 0.333333 | 0.666667 | 0.00000 | 0.166667 | 0.0.90492 |
| | Si(1) | 2c | 0.333333 | 0.666667 | 0.00000 | 0.166667 | 0.09508 |
| | P(2) | 1b | 0.000000 | 0.000000 | 0.50000 | 0.083333 | 0.90492 |
| | Si(2) | 1b | 0.000000 | 0.000000 | 0.50000 | 0.083333 | 0.09508 |

| Reliability factors | $R_1$ | $wR_2$ | $R_{int}$ | GooF |
|---|---|---|---|---|
| | 0.0254 | 0.0532 | 0.0691 | 0.953 |